\documentclass[a4paper, conference]{IEEEtran}
\IEEEoverridecommandlockouts
\usepackage{cite}
\usepackage{amsmath,amssymb,amsfonts}
\usepackage{algorithmic}
\usepackage{graphicx}
\usepackage{textcomp}
\usepackage{xcolor}
\usepackage{acronym}
\usepackage{soul} % Import the soul package
\usepackage{multirow} % for merging cells in rows
\usepackage{caption} % for customizing table captions
\usepackage{booktabs}
\usepackage{glossaries}
\usepackage{url}
\usepackage{hyperref}
\usepackage{comment}
\hypersetup{bookmarks=false}
\def\BibTeX{{\rm B\kern-.05em{\sc i\kern-.025em b}\kern-.08em
    T\kern-.1667em\lower.7ex\hbox{E}\kern-.125emX}}

\newacro{iot}[IoT]{Internet of Things}
\newacro{c2}[C\&C]{Command-and-Control}
\newacro{dns}[DNS]{Domain Name System}
\newacro{tii}[TII]{Technology and Innovation Institute}
\newacro{ml}[ML]{Machine Learning}
\newacro{cic}[CIC]{Canadian Institute for Cybersecurity}
\newacro{pcap}[pcap]{Packet Capture}
\newacro{nas}[NAS]{Network Attached Storage} 
\newacro{mt}[MT]{Mutation Testing}
\newacro{se}[SE]{Software Engineering} 
\newacro{dl}[DL]{Deep Learning} 
\newacro{rf}[RF]{Random Forest} 
\newacro{dt}[DT]{Decision Tree} 
\newacro{mlp}[MLP]{Multi-Layer Perceptron} 
\newacro{ids}[IDS]{Intrusion Detection System}

\begin{document}

\title{Real-time Threat Detection Strategies for Resource-constrained Devices}

\author{
 \IEEEauthorblockN{Mounia Hamidouche}
 \IEEEauthorblockA{\textit{Technology Innovation Institute} \\
 United Arab Emirates \\
 mounia.hamidouche@tii.ae}
 \and
 \IEEEauthorblockN{Biniam Fisseha Demissie}
 \IEEEauthorblockA{\textit{Technology Innovation Institute} \\
 United Arab Emirates \\
 biniam.demissie@tii.ae}
 \and
 \IEEEauthorblockN{Bilel Cherif}
 \IEEEauthorblockA{\textit{Technology Innovation Institute} \\
 United Arab Emirates \\
 bilel.cherif@tii.ae}

}

\maketitle

\begin{abstract}
As more devices connect to the internet, it becomes crucial to address their limitations and basic security needs. While much research focuses on utilizing \ac{ml} and \ac{dl} to tackle security challenges, there is often a tendency to overlook the practicality and feasibility of implementing these methods in real-time settings. This oversight stems from the constrained processing power and memory of certain devices (IoT devices), as well as concerns about the generalizability of these approaches. Focusing on the detection of DNS-tunneling attacks in a router as a case study, we present an end-to-end process designed to effectively address these challenges. The process spans from developing a lightweight DNS-tunneling detection model to integrating it into a resource-constrained device for real-time detection. Through our experiments, we demonstrate that utilizing stateless features for training the ML model, along with features chosen to be independent of the network configuration, leads to highly accurate results. The deployment of this carefully crafted model, optimized for embedded devices across diverse environments, resulted in high DNS-tunneling attack detection with minimal latency. With this work, we aim to encourage solutions that strike a balance between theoretical advancements and the practical applicability of ML approaches in the ever-evolving landscape of device security.

\end{abstract}


\section{Introduction}
In our dynamic technological era, prioritizing security for every connected device is crucial \cite{bitag}. However, the limited computational power and memory of numerous interconnected devices prevent manufacturers from implementing advanced security measures, such as Intrusion Detection Systems (IDS) in routers or anti-malware protection in sensors. This constraint exposes these devices to various cyber threats.
Our study seeks to underscore the importance of effectively balancing security measures with resource constraints for interconnected networks to be resilient against malicious activities. Specifically, before creating a detection model for security solutions based on ML or DL, it is essential to understand the specific context in which the model will be utilized and the limitations imposed by the deployment environment. Several previous works propose ML or DL-based approaches for intrusion detection such as the work in \cite{hamidouche2023enhancing} which proposes an effective detection model that uses packet-to-image transformation and DL technique. However, implementing this approach in a real-time environment would be challenging considering the extensive computational resources required associated with the data processing which renders the approach infeasible. Many other approaches suggest various ML-based techniques with features selected based on the dataset under study \cite{aiello2019unsupervised, berg2019identifying, dns-mitre, wang2022krtunnel}. However, while ML models combined with these features achieve high accuracy due to their customization for a particular dataset, they often overlook the feasibility of implementation in resource-constrained devices. 
The study presented in \cite{altuncu2021deep} introduces a real-time solution but assesses its approach in an environment that differs from the intended deployment scenario (for instance, using a computer acting as a proxy instead of an actual router designed to manage a substantial volume of traffic). In a real-time environment where a volume of traffic is expected and the decision has to be made in less than one millisecond, proposing approaches that would take several milliseconds before decision-making would make the approach impractical. In this study, we aim to show the challenges involved in proposing an end-to-end solution by picking one attack as a use case, namely, a DNS-tunneling attack.

% We choose the DNS-tunneling detection problem as a use case, emphasizing our comprehensive examination of both high-throughput and low-throughput scenarios \cite{nadler-lowthroughput}. %High-throughput DNS tuning is a specific type of DNS tuning that focuses on establishing a fast and reliable channel for transmitting a large volume of data through the DNS infrastructure. Low-throughput data filtration is another specific type of DNS tuning that involves slowly transmitting small amounts of data over time through DNS queries.

Several researchers have explored the subject of DNS-tunneling using one or more open-source tunneling tools in their experiment~\cite{aiello-dns2tcp, aiello-Dns2tcp-2, das-dnscat, jingkun-3tools, saeed-3tools, dns-botnet, CHEN2021102095, altuncu2021deep, al2021hybrid, 9432279, 9661022,adiwal2023dns}. The authors adopt a methodology in which they generate network traffic with various tools, analyzing it, and proposing different techniques for detecting the tunneling attack. Since traditional rule-based detection techniques are not effective when an adversary mounts highly customizable attacks, researchers propose more advanced ML or DL-based approaches to mitigate this issue by introducing diverse sets of features tailored to specific datasets. The features and models are crafted to efficiently detect attacks within their respective datasets.  To the best of our knowledge, no research has focused on evaluating the suitability of these features and models in real-world scenarios for real-time detection, particularly when considering deployment on actual real-time devices with limited resources.

Inspired by the work of \cite{doshi2018machine}, it is essential to consider certain constraints when defining the features that represent the DNS data to manage computational overhead in the router:

\paragraph{Lightweight features}
To effectively identify attacks within a router, which has the responsibility of managing substantial volumes of traffic, it is imperative to develop a lightweight \ac{ml} model that can operate efficiently with restricted resources \cite{hao2009detecting}.
\paragraph{Network configuration agnostic}
Routers, as the key components for managing network traffic, frequently meet various network setups. However, the majority of datasets utilized by academics to train their ML models are produced through the usage of DNS-tunneling tools on private networks~\cite{aiello2019unsupervised, berg2019identifying, dns-mitre}. Consequently, the suggested ML methods and features are predominantly tailored to this type of data produced by the tool and network. This inspires us to develop a detection model that can adapt to various network configurations and exhibit strong generalization skills.

\paragraph{High detection accuracy}
Our objective is to get high accuracy for a robust and trustworthy DNS-tunneling detection system within the router. In the context of security or anomaly detection, a high accuracy rate indicates that the system effectively distinguishes between normal and abnormal patterns, minimizing false positives and false negatives. 

\begin{figure}[!h]
\centering
\includegraphics[width=0.5\textwidth]{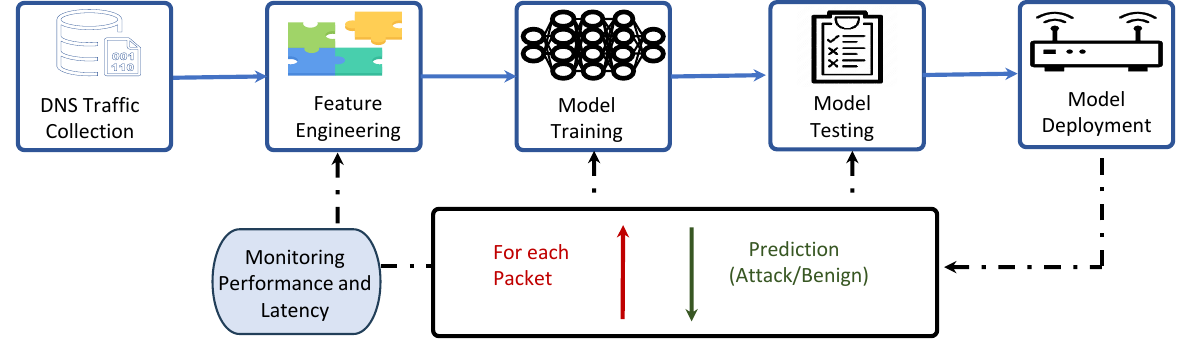}
\caption{End-to-End Pipeline: From Collection to Real-time Detection in the Router.}
\label{fig:pipeline}
\end{figure}

Based on these observations, we establish an end-to-end pipeline for real-time DNS-tunneling detection that considers the previously defined constraints. The process consists of several steps and iterations, as depicted in Fig.~\ref{fig:pipeline}.  The iterative process starts by collecting DNS traffic. Once data cleaning is performed, suitable features are identified that represent the data. We select a lightweight ML model that provides high detection accuracy, and we deploy the model on a specific router where we implemented an in-house IDS/IPS (Intrusion Detection System/ Intrusion Protection System) (details in Section~\ref{sec:experimental}). We then perform the attack and evaluate the performance of the IDS/IPS on the target device, in terms of detection accuracy and latency. We then improve the model by going back to the feature selection step until an acceptable trade-off between the accuracy and latency is reached, and the model generalizes in different network environments.

\section{Background and Related work}
This section gives background on DNS protocols, DNS-tunneling, and detection techniques. It also explores recent research on deployment considerations.

\subsection{Domain Name System (DNS)} 
DNS plays a pivotal role in facilitating communication between applications by using names as identifiers. Its primary function is to translate user-friendly domain names, such as "example.com," into corresponding IP addresses like "192.168.0.10." When a device initiates a DNS query for a domain name, the process typically begins with its local DNS resolver. If the resolver lacks the required information in its cache, it proceeds to query other DNS servers hierarchically, starting from the root servers, then progressing to the top-level domain (TLD) servers, and ultimately reaching authoritative name servers that store the specific domain's information. Despite its efficiency, the distributed and hierarchical nature of DNS renders it susceptible to attacks, such as DNS-tunneling.

\subsection{DNS-tunneling}
DNS-tunneling is a technique employed by attackers to transport non-DNS traffic by encapsulating it within DNS queries or responses. By exploiting the hierarchical structure of the DNS system, the objective is to facilitate covert communication and streamline data exfiltration \cite{bromberger2011dns}. In this process, the authoritative server is manipulated by the attacker, turning it into a server under their control. DNS messages are typically short, and responses are uncorrelated, meaning they may not arrive in the same order as the corresponding requests \cite{mockapetris1987domain}. 

An attacker might use \textit{high-throughput} DNS-tunneling or \textit{low-throughput} data exfiltration depending on the goal. \textit{High-throughput} DNS-tunneling involves establishing a bi-directional communication to transmit large volumes of data through the DNS infrastructure. This communication is utilized for data exfiltration, command-and-control (C2) operations. An example includes using a private VPN to bypass captive gateways that charge for Wi-Fi access, as seen in in-flight Internet scenarios or the OilRig threat group that utilizes covert DNS-tunneling techniques to steal data from organizations~\cite{OilRig}. On the other hand, \textit{low-throughput} data exfiltration focuses on transmitting small amounts of data over time through DNS queries to evade detection. This method aims to exfiltrate data discreetly, employing techniques such as compressions, encodings, encryption, and spreading data across multiple queries. An illustrative example involves the utilization of point-of-sale (POS) malware that employs DNS to illicitly acquire credit card details, disguising the communication as ping requests and DNS queries \cite{FrameworkPOS}. Additionally, instances of malware attributed to the OilRig threat group serve as another illustrative example~\cite{OilRig}.  In this work, we gather both \textit{high-throughput} and \textit{low-throughput} data.

\subsection{Resource-constrained devices}
Network devices may have limited memory and processing power, imposing constraints on the techniques used for anomaly detection, particularly in IoT networks \cite{kornaros2022hardware}. As we intend to deploy our ML model in a router, we encounter memory constraints that limit its ability to maintain extensive states. The introduction of caching, while an option, introduces latency and complexity. Therefore, an ideal algorithm for routers is either stateless or necessitates the storage of flow information within short time windows \cite{sivanathan2016low}.

\subsection{DNS-tunneling detection}

Network attack detection in general aims to identify patterns in the network traffic that deviate from normal behaviors. Effectively addressing DNS-tunneling attacks requires detection solutions targeting both high-throughput and low-throughput tunneling attempts. Traditional methods, such as broadly blocking DNS traffic or relying solely on Data Loss Prevention (DLP) tools \cite{liu2010data}, prove ineffective in detecting these sophisticated techniques. Rule-based approaches are also inadequate, as manual intervention becomes impractical when dealing with the substantial volume and frequent repetition of DNS requests. As previously highlighted, several researchers have explored the subject of DNS-tunneling attack detection using either ML or DL for their effectiveness~\cite{aiello2019unsupervised, berg2019identifying, dns-mitre, wang2022krtunnel, CHEN2021102095, Mahdavifar2021, alenzi, Zhan2022}. Specifically, these studies have focused on discovering the most effective features to input into the \ac{ml} model for detecting DNS-tunneling. Among these features, both \textit{stateless} and \textit{stateful} (temporal) features are created to represent each packet, serving as input for a \ac{ml} model. Stateless features primarily consist of packet header fields, while stateful features encompass aggregate flow information over specific time windows. In \cite{mahdavifar2021lightweight}, a list of commonly used stateless and stateful features for DNS-tunneling detection is provided. However, there is not enough focus on the practical side, especially when considering how feasible the detection using these features is on devices with limited resources. 
 % and related work

\section{Threat model}
\begin{figure*}[t] % Use [t] for top placement and [b] for bottom placement
  \centering
  \resizebox{!}{1.7in}{\includegraphics{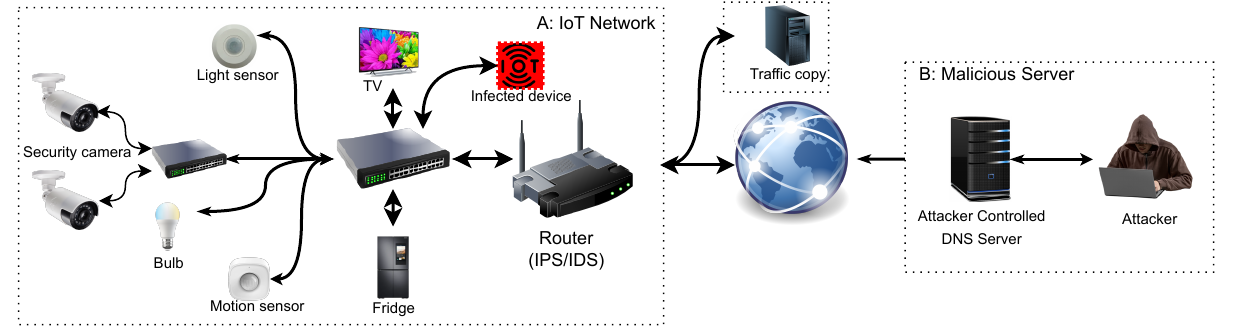}}
\caption{A Smart-home Network with an Infected Device (A) and Attacker Controlled Server (B).}
  \label{fig:network-diag}
\end{figure*}

Our model operates under the assumption that the targeted system is an IoT smart home as shown in Fig.~\ref{fig:network-diag}(A). In this network, we have two security cameras, a smart bulb, a light sensor, a smart fridge, a smart TV and a generic Linux-based infected device. The attacker's goals encompass both high- and low-throughput DNS-tunneling. The attacker could potentially compromise one or more IoT devices within the given network using lateral movements. Fig.~\ref{fig:network-diag}(B) shows the malicious attacker controlled DNS server. %Remote Code Execution (RCE), malicious firmware, or payload delivery. 
In the context of this network configuration, it is reasonable to infer that the DNS port remains open without any applied filtering, allowing unrestricted access. Moreover, our assumption includes the possibility that the attacker may have access to multiple domains, and some requests may occur non-consecutively, involving single UDP client-to-server conversations. Additionally, we consider that the user's router is configured to pass through a primary DNS that supports recursive resolution. Our goal is to deploy a detection model in the router to block tunneling attempts by infected network devices with a reasonable latency.

\section{Real-time DNS-tunneling detection pipeline}

In this section,  we elaborate on the end-to-end pipeline for DNS-tunneling detection, spanning from data collection to real-time detection within the router. 

\subsection{Network Traffic Collection}

We employ the publicly available tool called DNS2TCP available on GitHub~\cite{dns2tcp} to initiate the tunneling attack from the malicious node within our smart-home network to a remote DNS server hosted on a Virtual Private Server (VPS). This tool is specifically designed to encapsulate TCP traffic into various types of DNS queries, featuring a notable capability for high-throughput achieved by reducing packet size. The malicious node, integrated into the network, functions as an infected device responsible for either exfiltrating (sending out) or infiltrating (receiving) a small, but significant amount of data aligned with the sensors' state or command size. 
Sensor states such as the ON or OFF states leaking to a remote attacker could potentially pose security/privacy risks to the user/owner. Fig.~\ref{fig:network-diag} illustrates the implemented data exfiltration/infiltration setup in our IoT network. The tunneling node (the infected device) operates with two lists of random data:
\begin{itemize}
    \item The initial list models low-throughput communication with small data, such as sensor states.
    \item The second list simulates the high-throughput communication, such as data exfiltration.
\end{itemize}

To simulate attacker-sent commands, we utilized a random word generator (minimum length 1 letter), assuming that the generated commands align in size with the content exchanged covertly between a server and a client. The size alignment implies a correlation between the length of the commands and the concealed data exchanged. Once generated, the instructed commands can prompt the tunneling node to execute specific actions or initiate data exfiltration. This behavior is similar to what is observed in OilRig~\cite{OilRig} threat group's DNS-tunneling implementation. 

Using our lab setup, we passively collect both benign DNS requests and attack traffic over a 24-hour period, comprising high-throughput and low-throughput traffic, each spanning 12 hours. To ensure realistic conditions, no artificial interactions were imposed on the devices during the data collection; they were left to operate normally. The collected training data was stored in Packet Capture (PCAP) format using our experiment data collection platform (Traffic copy storage server in Fig.~\ref{fig:network-diag}). Table \ref{tab:tab2} provides details on the varied collection duration for each scenario and the number of collected packets. The DNS-tunneling dataset can be downloaded from~\url{https://bit.ly/DNS-Tunneling-dataset}.

  \begin{table}[t]
    \begin{center}
    \caption{Data Collection Duration and Packet Count.}
    \label{tab:tab2}
    \begin{tabular}{ l|ccc|ccc} 
    \toprule
    Scenario  & Duration & Packet Count \\
    \midrule
    \multicolumn{3}{c}{\bf{IoT Network}} \\
    Benign  &  12 hours & 38185\\
    % Benign & Low DNS traffic & 12 hours& \hl{12 hours} \\
    Attack &  12 hours & 31771 \\
    % Infected &  12 hours & 12 hours \\
    \midrule
    \multicolumn{3}{c}{{\bf \em{NEW\_ENV}}} \\
    Benign &  12 hours & 765 \\
    Attack & 12 hours & 17333 \\
    \bottomrule
    \end{tabular}
    \end{center}
    \end{table}

\subsection{Feature Selection}
\label{sub:feature selection}

To deploy the ML-based model in the router with limited resources, we focus on extracting network traffic features with the following characteristics:

\paragraph{Stateless} Stateless features typically involve only the analysis of packet header fields. Being independent of previous states, the model will require less computational power for their analysis. This is advantageous for devices with limited processing capabilities, as it minimizes the computational power during real-time detection. 

\paragraph{Network-configuration-agnostic}
A network configuration-agnostic approach implies that our ML-based detection model does not rely on features that capture specific network settings or configurations. This adaptability is beneficial as it allows operating effectively across diverse network environments without any further configuration of the model. 

\paragraph{Response-agnostic}
Focusing solely on analyzing DNS queries (disregarding responses) benefits implementing the model in a resource-constrained router. 

Considering the three constraints at hand, the selected features to represent our DNS data are given in Table \ref{tab:features}.

\begin{table*}[t]
  \centering
  \caption{Stateless Features.}
  \label{tab:features}
  \begin{tabular}{l|p{13cm}}
 % {>{\RaggedRight}p{4cm}|>{\RaggedRight}p{12cm}}
    \toprule
    Feature & Definition \\
    \midrule
    IP\_length & Refers to the total length of an Internet Protocol (IP) packet, including both the IP header and the payload. %In networking, an IP packet consists of two main parts: the header and the payload. The header contains essential information for routing and managing the packet, such as source and destination addresses, protocol information, and other control parameters. The payload, on the other hand, carries the actual data being transmitted. This measurement is expressed in octets or bytes and represents the entire size of the packet as it travels through a network.
    \\
    \midrule
    Query\_length & This metric is quantified in terms of the total number of bytes required to represent the entire query. \\
    \midrule
    Subdomain\_count/average & The count of labels in a domain, exemplified by the query name "www.scholar.google.com" consists of four labels separated by dots. The average length of labels in a subdomain. Plus the average length of labels in a subdomain. \\
    \midrule
    Shannon\_entropy & Calculated on the full query, including dots, denoted as %$H(X) = - \sum_{i=1}^{n} p(x_i) \cdot \log_2(p(x_i))$, 
    where $n$ is the number of distinct labels in the query and $p(x_i)$ is the probability of occurrence of label $x_i$. \\
    \midrule
    Max\_subdomain\_length & The number of bytes representing the longest subdomain. \\
    \midrule
    Query\_type & In DNS, the “query type” denotes the specific information requested about a domain. Various query types include A (IPv4 address), AAAA (IPv6 address), MX (mail servers), CNAME (canonical name alias), PTR (reverse lookup), NS (name servers), and SOA (start of authority for zone information). \\
    \midrule
    Special\_characters & The count of special characters such as dash (-), underscore (\_). \\
    \midrule
    Character\_frequency\_mean & Calculates the mean (average) of character frequencies in the DNS query. The formula is given by $CharFreqMean$ = $\frac{\sum{} f_i}{n}$, 
    where $f_i$ represents the frequency of the $i^{th}$ unique character, and $n$  is the number of unique characters in the query. \\
    \midrule
    Character\_frequency\_std & Computes the standard deviation of character frequencies in the DNS query. The formula is $ CharFreqStd = \sqrt{\frac{\sum{}(f_i - CharFreqMean)^2}{n}}$, 
    where $f_i$  is the frequency of the  $i^{th}$  unique character, $CharFreqMean$ is the mean character frequency, and $n$ is the number of unique characters in the query. \\
    \midrule
    Character\_frequency\_min/max & Identifies the minimum/maximum frequency of characters in the DNS query. This is determined by finding the smallest/longest value among the frequencies of unique characters. \\
    \midrule
    Numeric/Lower/ Upper\_characters & The count of numerical/lower/upper characters. \\
    \midrule
    Query\_class & Classification of a query based on its purpose, such as informational, navigational, or transactional. \\
    \bottomrule
  \end{tabular}
\end{table*}

\section{Experimental Analysis \& Evaluation}
\label{sec:experimental}

In this section, we evaluate the effectiveness of our approach with some experiments. The investigation is guided by the following research questions:

\begin{itemize}
    \item $RQ_1$: Can we develop a lightweight ML-based model with high accuracy in a controlled environment?
    \item $RQ_2$: How does the performance of the IDS/IPS change further to a real-time detection deployment in the router? 
    \item $RQ_3$: How does the performance of the IDS/IPS change when installed in a new environment?
\end{itemize}

The first research question $RQ_1$ investigates if we can obtain a lightweight model that can detect DNS-tunneling with high accuracy in a controlled environment. $RQ_2$ investigates the performance of the proposed IDS/IPS that uses the trained model for real-time detection. We investigate whether we achieve the same or acceptable detection accuracy without significantly degrading network latency within the smart-home environment shown in Fig.~\ref{fig:network-diag}. Lastly, $RQ_3$ delves into the performance assessment of the IDS/IPS in a novel network environment termed as {\em NEW\_ENV}, a setting for which the model did not undergo training. This examination is crucial for gauging the generalizability of the approach.

\subsection{$RQ_1$: Can we develop a lightweight model with high accuracy in a controlled environment?}

\subsubsection{Feature Importance}
In addition to using only stateless features, we analyze features' importance in a controlled environment, aiming to reduce their number.

\begin{figure}[ht]
    \centering
    \includegraphics[width=\linewidth]{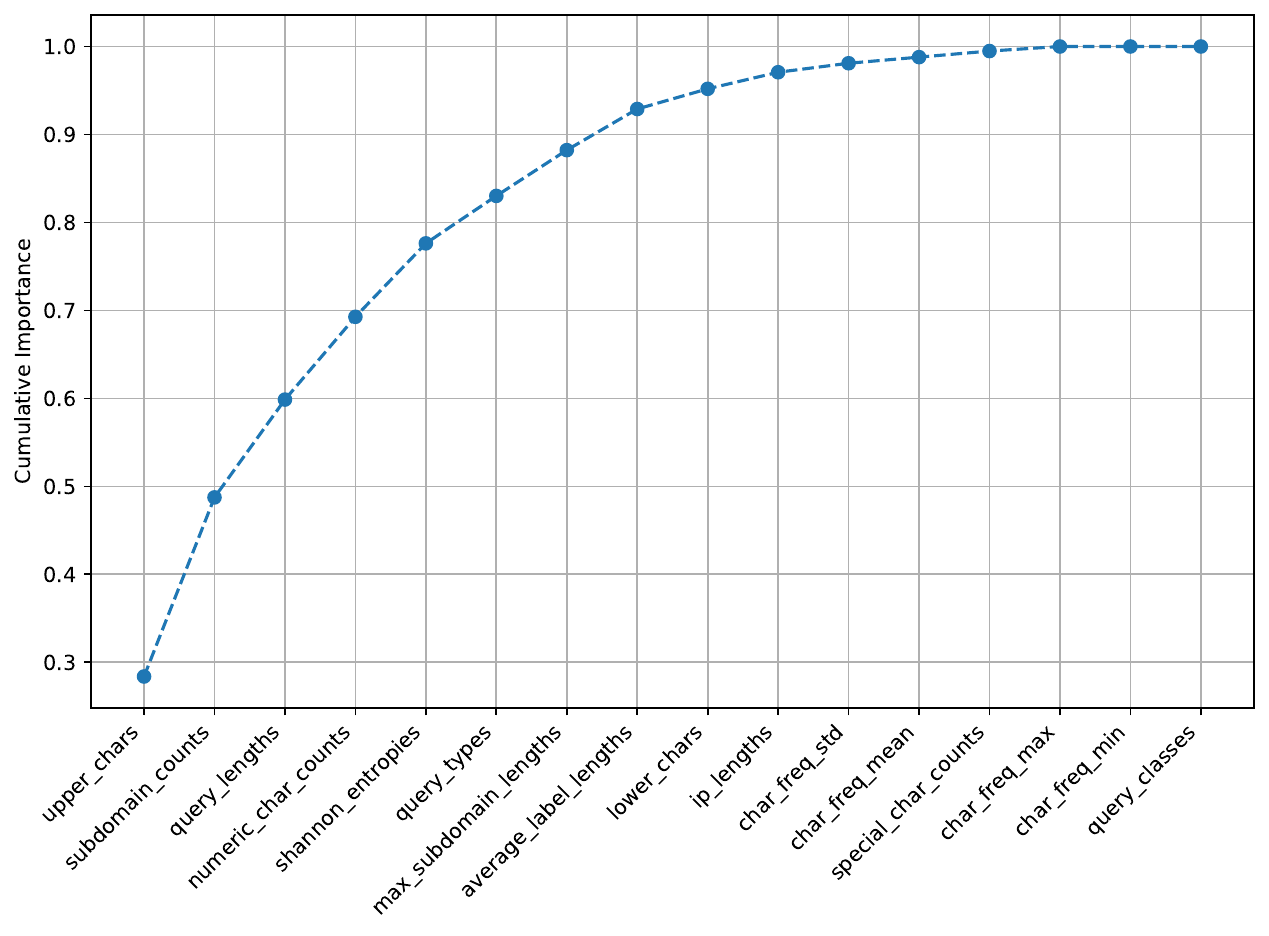}
    \caption{Cumulative Importance of Features.}
    \label{fig:cumulative_importance}
\end{figure}

\begin{figure*}[ht]
    \centering
    \includegraphics[width=14.5cm]{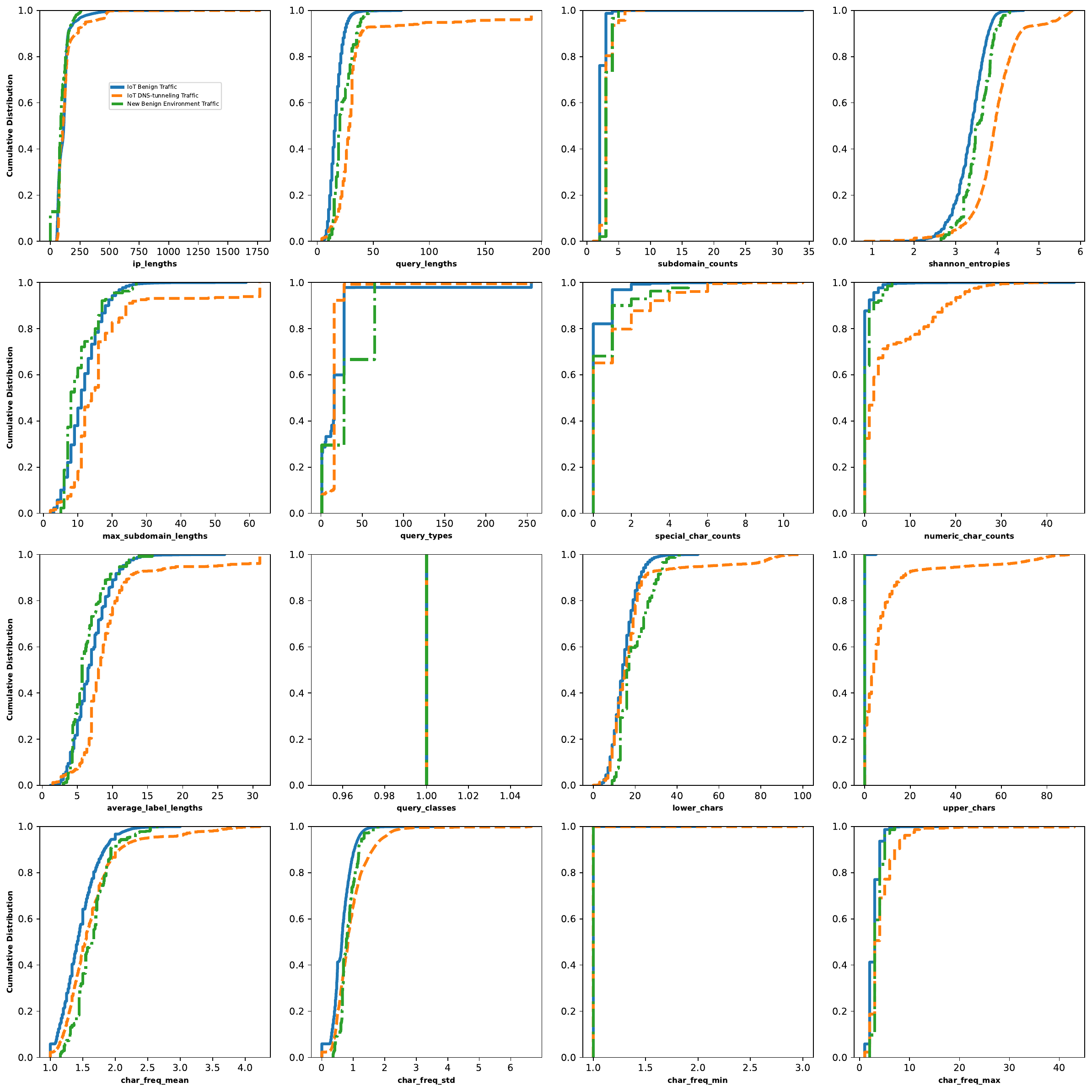}
    \caption{Features Distribution: Benign and Malicious IoT traffic VS. {\em NEW\_ENV} Benign traffic.}
    \label{fig:combined_cdf_plots}
\end{figure*}

In Fig. \ref{fig:combined_cdf_plots}, we illustrate the significance of individual stateless features related to DNS queries (disregarding responses) in predicting the target variable using random forest (RF). The y-axis denotes the feature importance scores, while the x-axis represents specific features considered by the RF classifier. Notably, the feature \texttt{Query\_classes} and \texttt{Character\_freq\_min} emerge with the highest importance as both attack and normal instances predominantly exhibit a constant value of 1, indicating its substantial influence on the model's predictions. Additionally, the feature \texttt{Character\_freq\_max} contribute significantly to the decision-making process. In addition, \texttt{Upper\_characters} and \texttt{Subdomain\_counts}, appear to be a less informative feature for DNS-tunneling detection, indicating limited discriminatory power. The feature \texttt{Query\_lengths} can also be disregarded as its marginal impact on distinguishing between benign and attack instances becomes evident as shown in Fig. \ref{fig:cumulative_importance}. Consequently, we are left with a total of 13 features among the initial 16 features.

\subsubsection{Binary Classification} We utilize decision trees, RF, k-nearest neighbors, support vector machines, and deep neural networks to accurately differentiate between DNS-tunneling attacks and legitimate traffic. This is achieved by selecting only the lightweight features shown in subsection \ref{sub:feature selection}. We implemented these models using the Scikit-learn Python library. All hyper-parameters were the default values. We trained the classifiers on a training set with 85\% of the combined normal and attack traffic and calculated classification accuracy on a test set of the remaining traffic.

\begin{table}[h]
\scriptsize
\centering
\caption{DNS-Tunneling Detection Accuracy: ML Models and Stateless Features.}
\begin{tabular}{c|c|c|c|c}
\toprule
\textbf{Model} & \textbf{ACC (\%)} & \textbf{REC (\%)} & \textbf{PRE (\%)} & \textbf{F1 (\%)} \\
\midrule
\textbf{RF} & \textbf{94.6} & \textbf{95.4} & \textbf{92.6} & \textbf{94}\\

\textbf{Decision Tree} & 93.2 & 92.3 & 92,7 & 92.5 \\

\textbf{K-Nearest Neighbors} & \textbf{95.1}
 & \textbf{96.8} & \textbf{92.2} & \textbf{94.4}\\

\textbf{Support Vector Machine} & 91.2 & 95.2 & 84.9 & 89.8 \\

\textbf{Deep Neural Network} & 93.8 & 97.5 & 88.7 & 92,9 \\
\bottomrule
\end{tabular}
\label{Table:classification_attack}
\end{table}

We use four metrics for the performance evaluation: accuracy (ACC), precision (PRE), recall (REC), and F1-score (F1). We use accuracy to evaluate the overall performance of a classifier. Precision, recall, and F1-scores to assess the performance of every traffic class. 

\begin{equation*}
    \text{ACC} = \frac{\text{TP+TN}}{\text{TP+FP+FN+TN}}, \ \ \ \text{PRE} = \frac{\text{TP}}{\text{TP+FP}},
\end{equation*}
\begin{equation*}
    \text{REC} = \frac{\text{TP}}{\text{TP+FN}}, \ \ \ \ \text{F1-score} = \frac{2. \text{ PRE . REC}}{\text{PRE + REC}},
\end{equation*}
Whereas true positive (TP) is the number of instances correctly classified as attack, true negative (TN) is the number of instances correctly classified as benign, false positive (FP) is the number of instances incorrectly classified as attack, and false negative (FN) is the number of instances incorrectly classified as benign.

Observing Table \ref{Table:classification_attack}, it is evident that both the RF and K-nearest neighbors exhibit the most favorable outcomes. We opted for RF over KNN because we were already integrating RF into the router, and creating a custom model wrapper for each new model would be time-consuming. 

\subsection{$RQ_2$: How does the performance of the IDS/IPS change when deploying the lightweight model on a router in a real-time environment?}

\subsubsection{Router specification}
We deployed the lightweight model in an IDS/IPS environment within a router, specifically using the open-source Turris Omnia~\cite{turris}. The router features a 1.6GHz dual-core ARM CPU, 2GB RAM, and runs TurrisOS, a fork of OpenWRT. Its open-source, customizable nature, and design for home/small office use make it an ideal candidate for our experiment.

\subsubsection{Real-time traffic capture}
To capture real-time traffic, we relied on Netfilter~\cite{netfilter-wiki} from the Linux subsystem. Netfilter constitutes a framework embedded in the Linux kernel, offering diverse hooks that enable kernel modules to enroll callback functions for events associated with network packets as they navigate through the Linux networking stack. Serving as an integral element of the Linux firewall and packet filtering infrastructure, Netfilter is primarily employed for the establishment of packet filtering rules.

\begin{figure}[ht]
    \centering
    \includegraphics[trim={1cm 0.3pt 0.3pt 15pt,clip},width=1.03\linewidth]{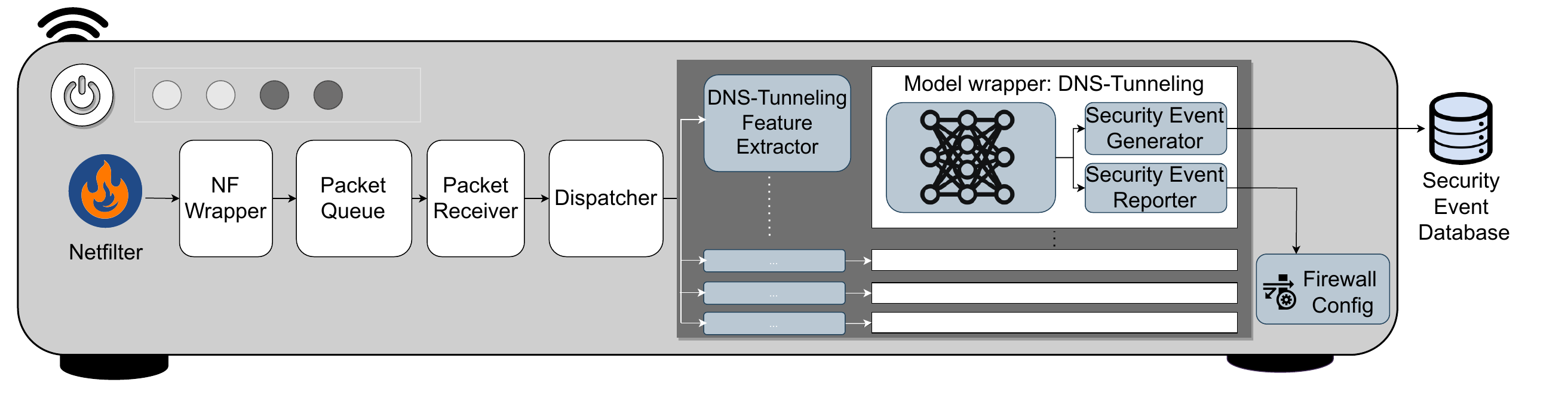}
    \caption{Real-time Detection on the Router.}
    \label{fig:ondevice-detection}
\end{figure}

We use the Netfilter queue to pass incoming packets from the kernel to a userspace for further processing and decision-making (verdict). We use \texttt{iptables} to set up a rule to redirect incoming packets to the Netfilter queue. We implemented the userspace application in embedded C++ that processes the packets in the queue. The application then decides whether to allow or drop the packet based on the detection model and configures the firewall. If a packet is dropped, a security event is generated and the network administrator can review the event on a dashboard and revert the configuration if necessary. Below, we provide a brief description of the different components from real-time traffic capture to decision-making process within the router, see Fig.~\ref{fig:ondevice-detection}:

\begin{itemize}
    \item {\bf NF Wrapper} This is a wrapper for Netfilter queue implemented using \texttt{libnetfilter\_queue} that provides an interface to interact with the Netfilter queue framework in the Linux kernel.
    \item {\bf Packet Queue} Internal packet queue management.
    \item {\bf Packet Receiver} It dequeues a packet at a time from the internal queue and sends it to further processing.
    \item {\bf Dispatcher} It performs preliminary analysis on the packet (e.g., select the right protocol) and dispatches the packet to the right processor. In this case, DNS packets will be dispatched to the DNS-tunneling processor.
    \item {\bf Feature Extractor} Detecting various attacks may necessitate distinct features (e.g., DoS attack detection relying on \texttt{Source\_IP}, while DNS-tunneling detection requires \texttt{Query\_classes}). Extracting all features for every attack incurs unnecessary overhead. Our approach mitigates this by allowing inclusion of new attack detection with corresponding feature extractors. The dispatcher routes packets based on the attack specification (e.g., for DNS-tunneling attack, \texttt{protocol == DNS}). Rigorous testing is necessary to confirm the matching of feature values extracted from the Linux kernel and those from Wireshark during training phase (e.g., rounding of values). This is performed by uploading the training PCAP file to the router and using the Feature Extractor to extract the features. The test passes if the feature values match the values extracted within the training environment. Note that updating the ML model might require reimplementing the Feature Extractor to match the set of features. 
    % \hl{mention the testing of the feature values from the Python vs the C++ feature extractor match}
    \item {\bf Model Wrapper} We can have various model wrappers tailored for distinct attacks, receiving the previously extracted features and providing predictions (i.e., attack or benign). In this experiment, we implemented a single wrapper for the DNS-tunneling detection model. However, the evaluation of performance, comparing one model for all possible attacks versus employing one lightweight model per attack, is identified as future work. The wrapper is a C++ version of the model optimized for embedded systems. We use {\em emlearn}~\cite{emlearn} to generate a C++ header corresponding to the model. The generated model is way more optimized than the initial Python based model as it eliminates all the overhead of generic layers implementation. Compiling the model as a native binary eliminates the additional overhead of the Python interpreter. If an attack is detected, the packet is dropped, and the event is managed by the two subprocesses:
    \begin{itemize}
    \item {\bf Security Event Generator} It generates the right kinds of events to be sent to a database for visualization purposes (e.g., in a SIEM). This component is also attack-specific because the description, the attack source, and the attack destination could be specific to the attack.
    \item {\bf Security Event Reporter}: It generates the right security report that will be used by the firewall configuration to block further such attacks.
    \end{itemize}
    \item {\bf Firewall Config} When an attack is detected, the firewall is automatically configured based on the right security event report generated in the previous step.
\end{itemize}

\begin{table}[h]
\scriptsize
\centering
\caption{Assessing Trained RF Model in {\em NEW\_ENV}.}
\begin{tabular}{c|c|c|c|c}
\toprule
\textbf{Model} & \textbf{ACC (\%)} & \textbf{REC (\%)} & \textbf{PRE (\%)} & \textbf{F1 (\%)} \\
\midrule
\textbf{RF} & \textbf{93.05} & \textbf{92.23} & \textbf{99.92} & \textbf{95.92}\\
\bottomrule
\end{tabular}
\label{Table:classification_attack-router}
\end{table}

\begin{table}[h]
  \centering
  \caption{Router Latency for DNS-Tunneling Detection.}
  \begin{tabular}{lc}
    \toprule
    \textbf{One Packet Process} & \textbf{Latency (ms)} \\
    \midrule
    Feature Extraction, Detection \& Verdict & $<$1 \\
    \bottomrule
  \end{tabular}
  \label{tab:latency}
\end{table}

\subsubsection{Real-time Classification} The model's performance in the router closely aligns with the results presented in Table \ref{Table:classification_attack}, which is expected, given that the incoming data to the router is sourced from the same environment used for training our machine learning model.

\subsection{$RQ_3$: How does the performance of the IDS/IPS change when installed in {\em NEW\_ENV}?}

To answer this question, we set up a new network (called {\em NEW\_ENV}) that includes two Linux virtual machines, one Android phone and a host computer connected to the router (IDS/IPS). We used the same tunneling tool in order to perform the attack with both high- and low-throughput random data. 

While evaluating the trained RF model in {\em NEW\_ENV}, Table~\ref{Table:classification_attack-router} shows that the model achieved highly promising results, affirming its adaptability and robustness. The accuracy of 93.05\% underscores the model's capability to make correct predictions across diverse scenarios. Notably, a precision of 99.92\% indicates a minimal rate of false positives.

In terms of latency, as shown in Table \ref{tab:latency},  our investigation shows a remarkable efficiency in the router's packet processing, with each packet taking less than 1 ms for a complete cycle. This rapid processing involves the packet's entry into the router, feature extraction, detection of the DNS-tunneling attack and verdict. The latency of less than 1 ms attests to the router's adeptness in quick decision-making, crucial for timely identification and mitigation of security threats in dynamic network environments such as routers.

%%%%%%%%%%%%%%%%%%%%%%%%%%%%%%%%%%%%%%% Detection accuracy %%%%%%%%%%%%%%%%%%%%%%%%%%%%%%%%%%%

\section{Discussion and Future Work}

This study addresses challenges in providing practical security solutions for resource-constrained environments. We successfully detected DNS-tunneling attack traffic using ML-based classification algorithms and lightweight, stateless, and network configuration-agnostic features, in both controlled and real-time settings. Implementing the IDS/IPS in embedded C++ and optimizing the model for embedded systems minimized network overhead during detection.

Identifying potential features and advanced ML techniques is kept for future work to enhance anomaly detection in device networks. However, the feasibility of these strategies must be assessed in real-world environments. While a 100\% accurate solution is appealing, its practicality needs evaluation, considering potential delays in real-time decision-making.

The challenge of updating the ML model without restarting is an intriguing avenue for future investigation, particularly exploring retraining vs. incremental training. Furthermore, updating the embedded model is easier if features remain unchanged; however, if the set of features changes, reimplementation of the feature extractor on the device is necessary. This is due to the manual implementation of the packet header extractor interacting with the queuing buffer, as adapting the queuing logic from Linux kernel-space to user-space requires implementing the appropriate packet filters for each attack detection model (see Fig.~\ref{fig:ondevice-detection}).

While our proposed technique processes one packet at a time, assessing its performance under scenarios requiring packet buffering for decision-making (e.g., DoS attacks) would be intriguing. Finally, the evaluation of incorporating a model that detects more than one attack versus employing multiple models for distinct attacks is designated as future work.

\section{Conclusion}
Our study focuses on the vital connection between device networks security and limited resources, emphasizing the importance of discovering practical and customized solutions. In this work, we demonstrated that stateless features combined with simple ML models can accurately detect DNS-tunneling attacks in a realistic environment. Our choice of features is motivated by the fact that we deploy the model on a router, which has limited resource constraints. We tested various ML models on a dataset collected from an IoT network. We retained the RF model for its highest accuracy. Subsequently, deployed on the router, the model successfully detected the DNS-tunneling attack using real-time data from a virtual machine, showcasing its generalization capabilities.

\begin{comment}
1- Check to which research question, the accuracy table corresponds to ok. 2- Discuss how different the dataset in the new environment is from the one we have used for training the ML model. 3- Check the figures, what to change what to add. 4- Check how many packets there are for each of the low and high throughput. 5- Read again the whole paper to make it fit into 8 pages. 6- add that resposne-agnostic in the different parts of the work like intro ok.

\end{comment}

\bibliographystyle{plain}
\bibliography{bib.bib}

\end{document}